\documentclass[11pt,twoside]{book}
\usepackage{konkolyproc2}
\usepackage{longtable}
\usepackage{amsmath,amssymb}
\usepackage{graphicx}
\usepackage{lscape}
\usepackage{index}
\usepackage{natbib}
\usepackage{bigdelim}
\usepackage{multirow}
\makeindex

\begin{document}

\pagestyle{myheadings}
\setcounter{equation}{0}\setcounter{figure}{0}\setcounter{footnote}{0}\setcounter{section}{0}\setcounter{table}{0}\setcounter{page}{1}
\markboth{Netzel, Smolec \& Moskalik}{RRL2015 Conf. Papers}
\title{Nonradial modes in RR Lyrae stars from the OGLE Collection of Variable Stars}
\author{Henryka Netzel$^1$, Rados\l aw Smolec$^2$ \& Pawe\l~Moskalik$^2$ }
\affil{$^1$Warsaw University Astronomical Observatory, Warsaw, Poland\\
$^2$Nicolaus Copernicus Astronomical Center, Warsaw, Poland}

\begin{abstract}
The Optical Gravitational Lensing Experiment (OGLE) is a great source of top-quality photometry of classical pulsators. Collection of variable stars from the fourth part of the project contains more than 38 000 RR Lyrae stars. These stars pulsate mostly in the radial fundamental mode (RRab), in radial first overtone (RRc) or in both modes simultaneously (RRd). Analysis of the OGLE data allowed to detect additional non-radial modes in RRc and in RRd stars. We have found more than 260 double-mode stars with characteristic period ratio of the additional (shorter) period to first overtone period around 0.61, increasing the number of known stars of this type by factor of 10. Stars from the OGLE sample form three nearly parallel sequences in the Petersen diagram. Some stars show more than one non-radial mode simultaneously. These modes belong to different sequences.

\end{abstract}

\section{Introduction}
RR Lyrae stars are classical pulsating stars. They are known to pulsate mostly in radial fundamental mode (RRab) or in first overtone (RRc). Among RR Lyrae stars there are also double-mode pulsators which pulsate in fundamental mode and first overtone simultaneously (RRd, green asterisks in Fig.~\ref{fig.pet}) or in fundamental mode and second overtone (red triangles in Fig.~\ref{fig.pet}). The latter group was discovered mostly thanks to excellent space observations. Observations also revealed a group of RR Lyrae stars in which an additional non-radial mode is excited. These stars pulsate in a radial first overtone (RRc or RRd stars) and in an additional mode with shorter period, $P_X$. Period ratio of the additional mode to the first overtone is in the range of $0.60-0.64$ (blue circles in Fig.~\ref{fig.pet}). The most typical value is around 0.61. Such period ratio cannot correspond to two radial modes \citep{pamsm15}. Hence, the additional mode (0.61 mode in the following) must be non-radial. This type of pulsations was discovered in 23 RR Lyrae stars, based both on ground and on space observations \citep[for a summary see][]{pamsm15} and recently in 18 RR Lyrae stars from M3 \citep{jurcsik_M3}.

\section{Data Analysis}

This 0.61 mode has always very low amplitude in a milimagnitude regime. Almost all stars observed with space telescopes showed this non-radial mode \citep{chadid,molnar,pamsm15,szabo_corot}. Low amplitude of this mode makes it very difficult to detect it in the ground-based observations. We decided to search for the 0.61 mode in RR Lyrae stars of the Galactic bulge observed by the OGLE project \citep{ogle-iv}. Although quality of ground-based photometry is lower than of the space-based photometry, the number of observed stars is much higher in ground observations.

\begin{figure}[!ht]
\centering
\includegraphics[width=0.7\textwidth]{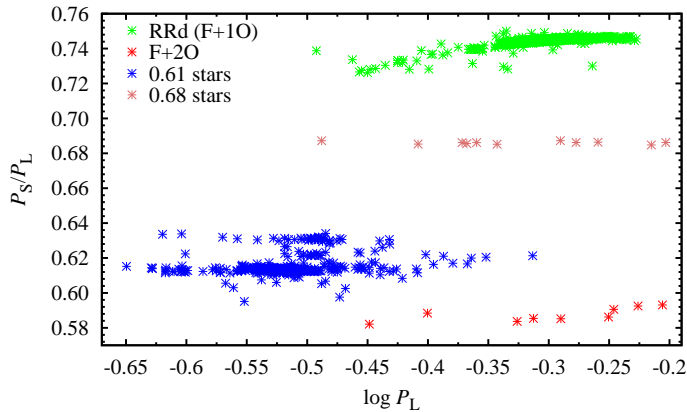}
\caption{Multi-mode pulsations of RR Lyrae stars in the Petersen diagram.} 
\label{fig.pet} 
\end{figure}

For the analysis we chose all stars available from the OGLE-III \citep{ogle-iii-bulge} pulsating in first overtone. Input sample consists of 4989 RRc stars and 91 RRd stars. RRc stars were analysed with an automatic method using dedicated software \citep[for details see][]{061oiii} and 91 RRd stars were analysed manually. It resulted in a detection of 147 stars with 0.61 mode ($3\%$ of the sample).

In OGLE-IV there are more than 10 000 RRc stars. Because our analysis was focused on a search for low amplitude signals, we decided to choose only the most frequently observed stars. These are located in OGLE fields 501 and 505 \citep[see position of observational fields in fig.~15 in][]{ogle-iv}. Input sample consists of 485 RRc stars, which were analysed manually. The rest of stars observed during the fourth phase will be analysed automatically. We detected 131 RRc stars, from which 115 are new discoveries \citep{061oiv}. With best quality data the 0.61 mode is found in $27\%$ of stars. We also detected another group of double-mode radial--non-radial RR Lyrae stars \citep[magenta crosses in Fig.~\ref{fig.pet},][]{068}.

\section{Results}

Petersen diagram for multi-mode RR Lyrae stars is presented in Fig.~\ref{fig.pet}. All known 0.61 stars are marked with blue circles. Thanks to the OGLE data we know altogether 303 0.61 stars, compared to 41 previously known. Which such numerous sample we can see for the first time three sequences in the Petersen diagram. The lowest sequence, around period ratio $0.61$, is most populated. The highest sequence is less populated and is located around period ratio $0.63$. Between these two, there is a third sequence, least populated, but well separated from the other two; it is clustered around period ratio $0.62$. Structures we detect in the power spectra of the stars are very broad \citep[see fig.~3 in][]{061oiv}. In the time-domain it corresponds to variability of amplitude and phase of the signal. This behaviour of 0.61 mode is confirmed by other studies both from ground-based \citep[e.g.][]{jurcsik_M3} and space-based observations \citep[e.g.][]{pamsm15,molnar,szabo_corot}. Most exciting result is a discovery of stars which have three signals in the power spectrum corresponding to three sequences in the Petersen diagram. Power spectra of 6 such star are presented in fig.~5 in \cite{061oiv}. 

\begin{figure*}
\centering
\resizebox{0.49\hsize}{!}{\includegraphics{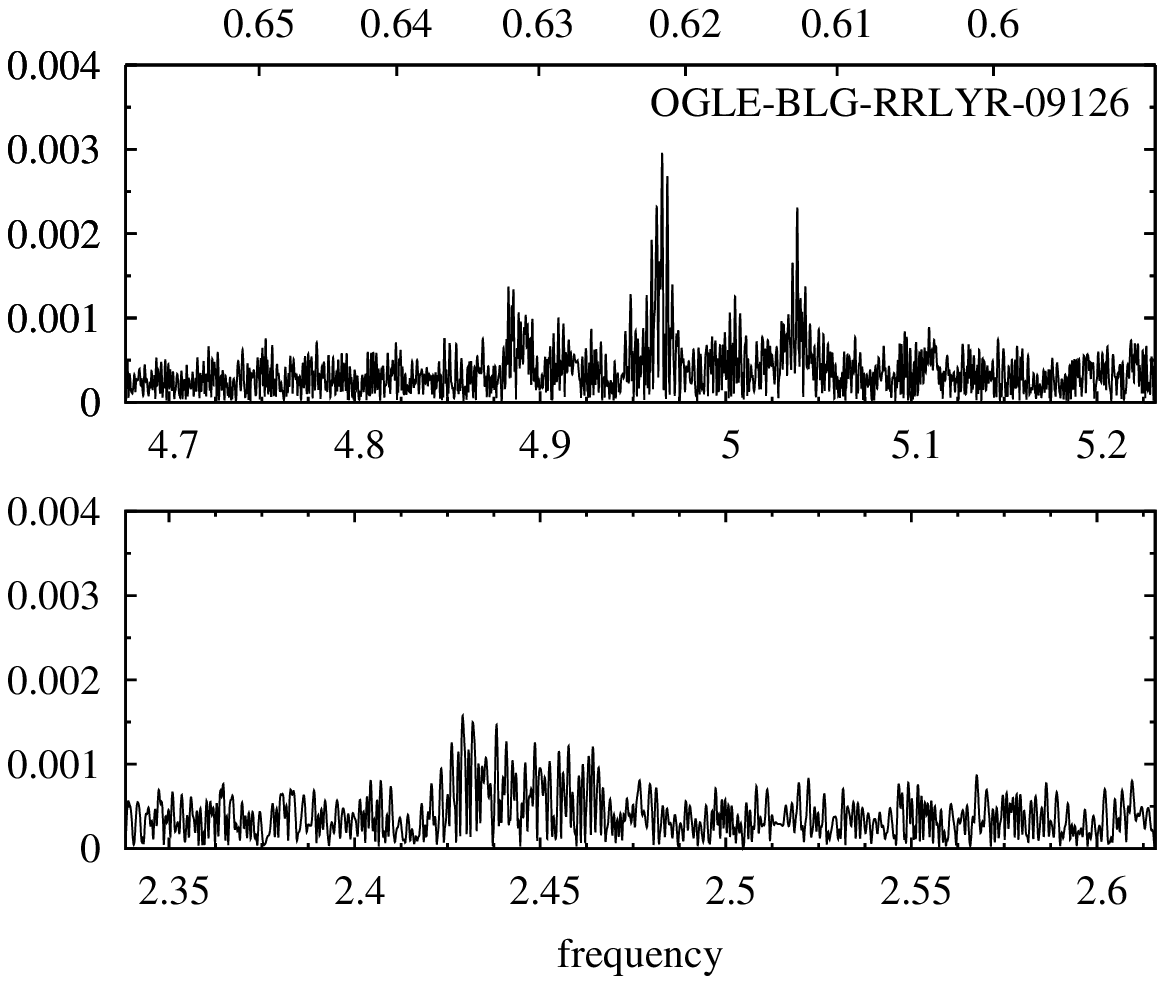}}
\resizebox{0.49\hsize}{!}{\includegraphics{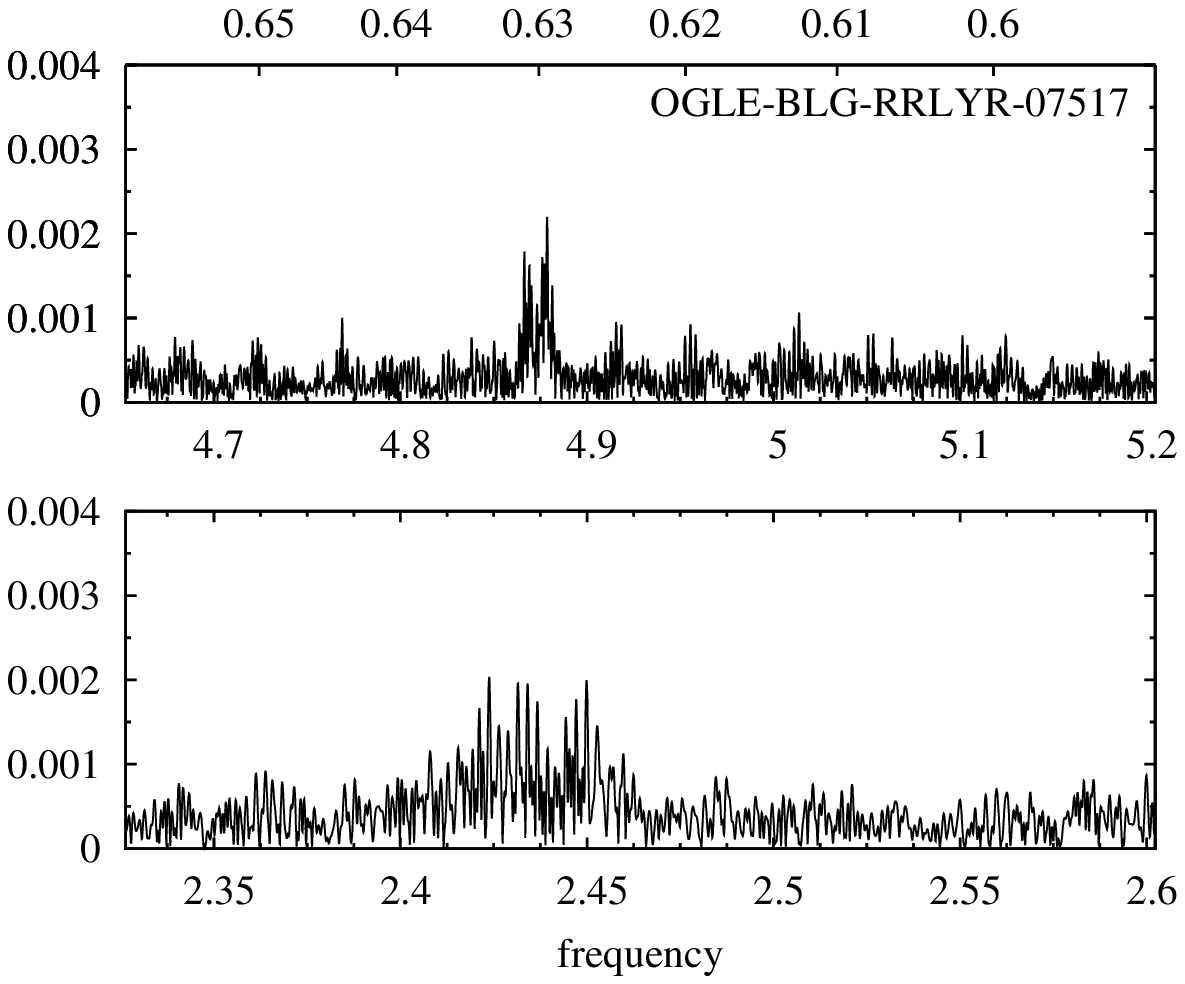}}\\
\vspace*{.2cm}
\resizebox{0.49\hsize}{!}{\includegraphics{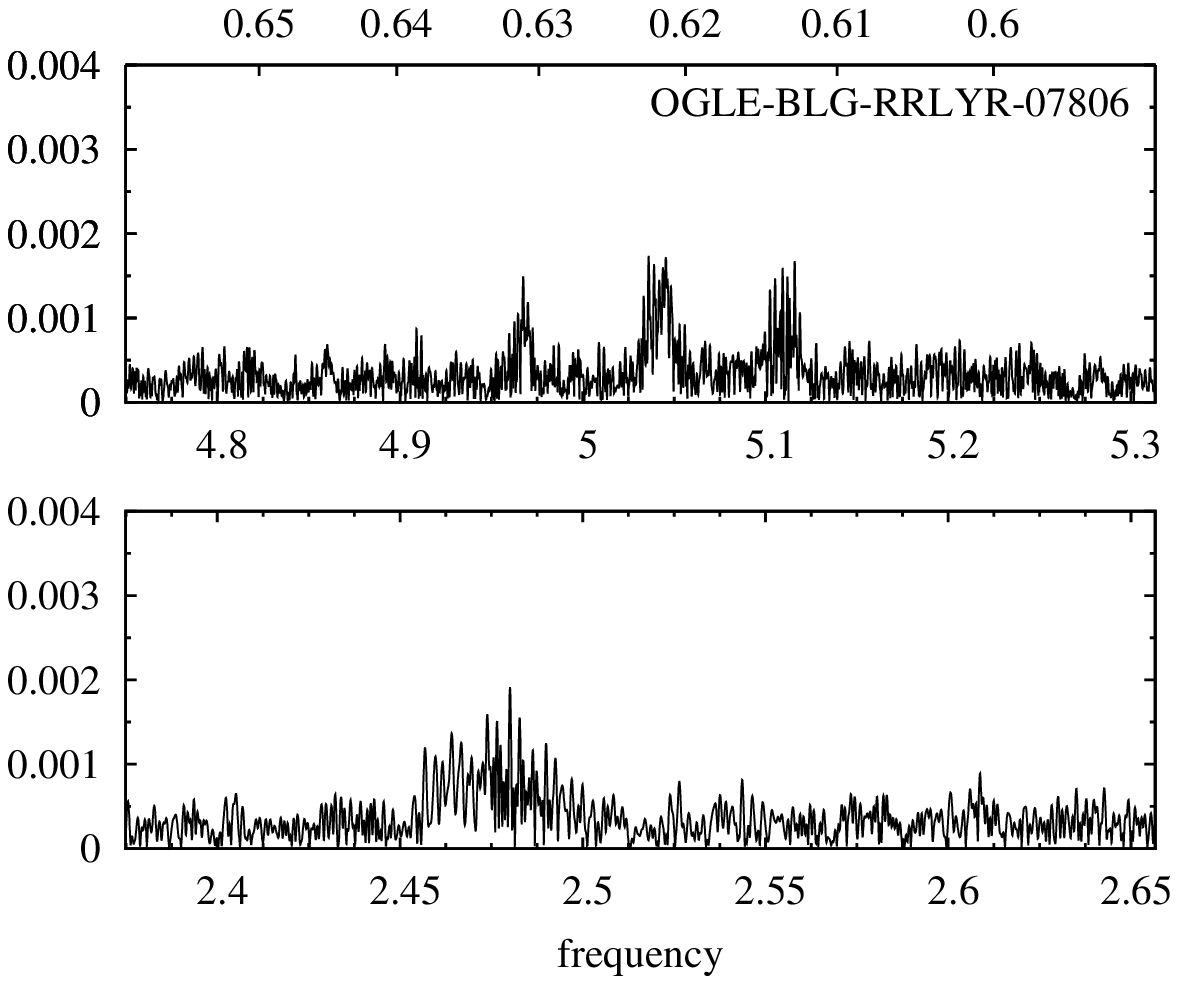}}
\resizebox{0.49\hsize}{!}{\includegraphics{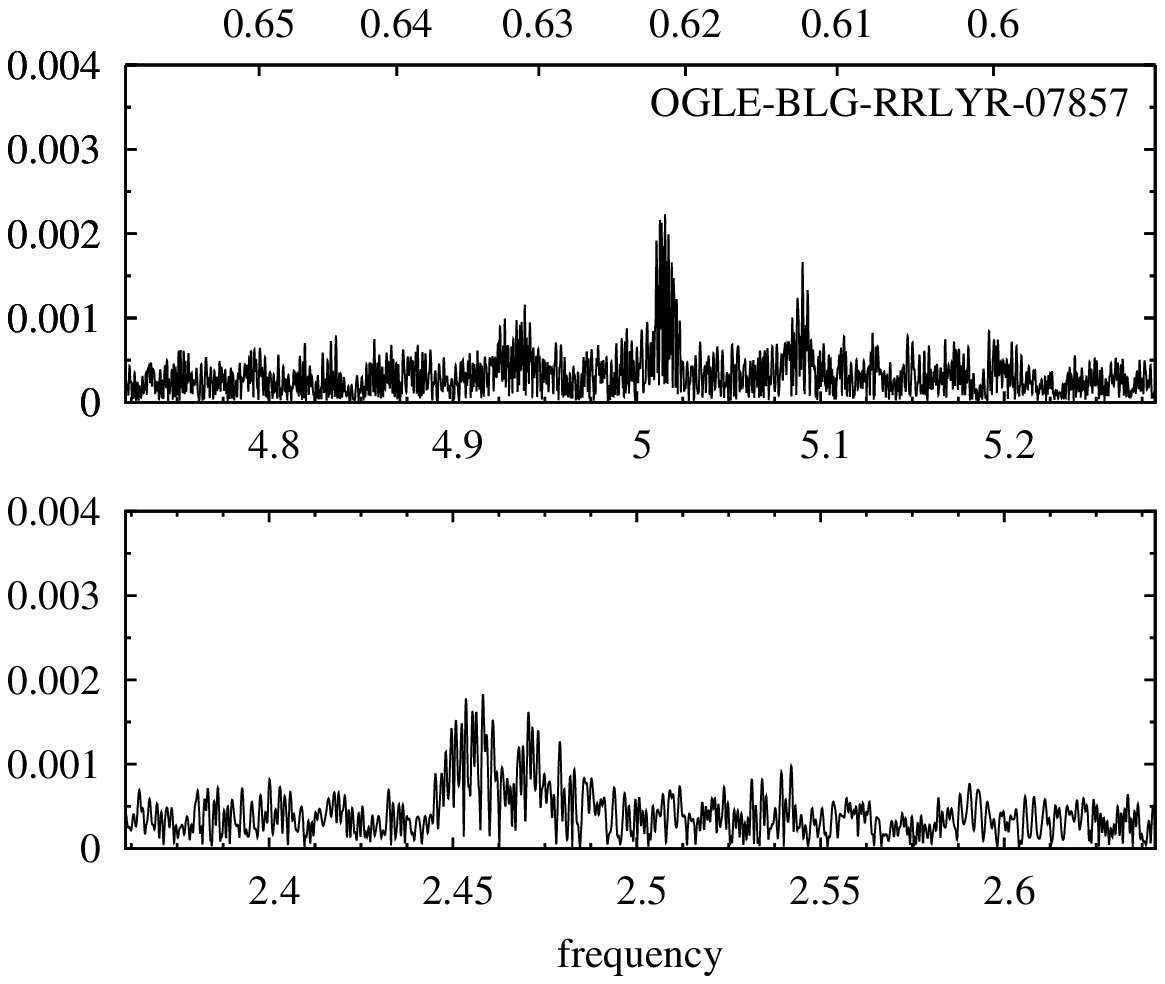}}\\
\caption{Frequency spectra centered at frequency range
         characteristic for additional mode (top panels) and its
         $1/2$ subharmonic (bottom panels) for a sample of 0.61
         stars. Directly underneath a signal with frequency $f$ (top panel) is
         located its subharmonic with frequency $1/2f$ (the bottom
         panel). Period ratio of the additional mode (top panels) to first overtone is indicated above top panel.}
\label{fig.subharm}
\end{figure*}

Another common property of 0.61 stars is occurence of subharmonics of the additional mode, both $1/2~f_X$ and $3/2~f_X$. We detected signal at subhamonic frequency in 26 stars, which constitute $20~\%$ of the OGLE-IV 0.61 stars. Additional modes and their subharmonic are shown in Fig.~\ref{fig.subharm} for selected stars \citep[see also fig.~11 in][]{061oiv}. Structures of subharmonics are very complex. They appear in power spectra as wide bands of excess power. Before our analysis of the OGLE data, subharmonics were detected only in space observations. Typically, presence of subharmonic frequency indicates period doubling of the parent mode \citep[see e.g.][]{doubling}. However, there is another possibility proposed by Dziembowski (these proceedings). Signal around $1/2~f_X$ can be a real mode present in a star, while signal at $f_X$ is its harmonic.

\begin{figure}[!ht]
\centering
\includegraphics[width=0.7\textwidth]{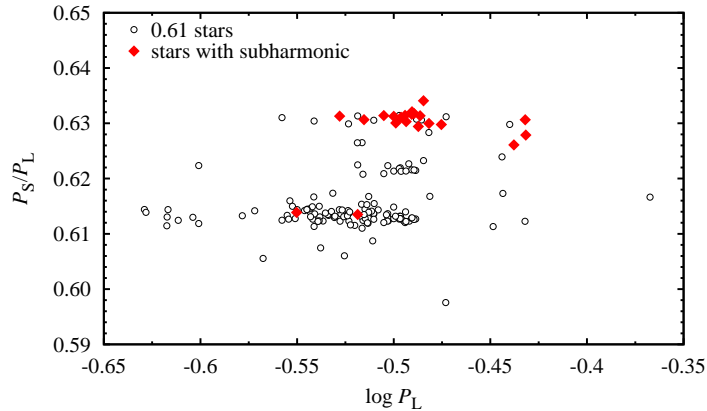}
\caption{Petersen diagram of 0.61 stars. Stars found in OGLE data are marked with black circles. Stars which have frequency at subharmonic are marked with red diamonds (only stars which show significant signal at $f_X$ {\bf and simultaneously} at around $1/2~f_X$ are marked).} 
\label{fig.sh} 
\end{figure}

The majority of signals at subharmonic frequency range, $75~\%$, correspond to 0.63 sequence. In Fig.~\ref{fig.sh} we present Petersen diagram for 0.61 stars from OGLE data. With red diamonds we marked stars for which signal at subharmonic is detected. Majority of these stars occupy the highest sequence. This finding supports the model proposed by Dziembowski (these proceedings).

\section*{Acknowledgements}
This research is supported by the Polish National Science Centre through grant DEC-2012/05/B/ST9/03932.

\end{document}